\documentclass[11pt,graybox]{svmult}

\usepackage[usenames,dvipsnames]{xcolor}%
\usepackage [colorlinks=true,urlcolor=blue,citecolor=blue,linkcolor=blue]{hyperref} %

\usepackage{mathptmx}       
\usepackage{helvet}         
\usepackage{courier}        
\usepackage{type1cm}        
\usepackage{makeidx}         
\usepackage{graphicx}        
\usepackage{multicol}        
\usepackage[bottom]{footmisc}

\usepackage{mathbbol,amssymb,latexsym,amsfonts,amsmath}

\usepackage[compress]{cite}
\usepackage{soul}
\usepackage{url}
\usepackage[utf8x]{inputenc}

\newcommand{\Pa}{\mathcal{P}}
\newcommand{\A}{\mathcal{A}}
\newcommand{\Dc}{\mathbf{X}}
\newcommand{\Ec}{\mathcal{E}}
\newcommand{\M}{\mathcal{M}}
\newcommand{\R}{\mathbb{R}}

\title*{Learning differential module networks across multiple experimental conditions}

\author{Pau Erola, Eric Bonnet and Tom Michoel}

\institute{Pau Erola \at Division of Genetics and Genomics, The Roslin Institute, The University of Edinburgh, Midlothian EH25 9RG, Scotland, United Kingdom \and Eric Bonnet \at Centre National de Recherche en Génomique Humaine, Institut de Biologie François Jacob, Direction de la Recherche Fondamentale, CEA, Evry, France \and Tom Michoel \at Division of Genetics and Genomics, The Roslin Institute, The University of Edinburgh, Midlothian EH25 9RG, Scotland, United Kingdom. Correspondence to \email{Tom.Michoel@roslin.ed.ac.uk}}

\begin{document}

\maketitle

\abstract{Module network inference is a statistical method to reconstruct gene regulatory networks, which uses probabilistic graphical models to learn modules of coregulated genes and their upstream regulatory programs from genome-wide gene expression and other omics data. Here we review the basic theory of module network inference, present protocols for common gene regulatory network reconstruction scenarios based on the Lemon-Tree software, and show, using human gene expression data, how the software can also be applied to learn differential module networks across multiple experimental conditions.}


\section{Introduction}
\label{sec:introduction}

Complex systems composed of a large number of interacting components often display a high level of modularity, where independently functioning units can be observed at multiple organizational scales \cite{newman2006b}. In biology, a module is viewed as a discrete entity composed of many types of molecules and whose function is separable from that of other modules \cite{hartwell1999}. The principle of modularity plays an essential role in understanding the structure, function and evolution of gene regulatory, metabolic, signaling and protein interaction networks \cite{qi2006}. It is therefore not surprising that functional modules also manifest themselves in genome-wide data. Indeed, from the very first studies examining genome-wide gene expression levels in yeast, it has been evident that clusters of coexpressed genes, i.e.\ sharing the same expression profile over time or across different experimental perturbations, reveal important information about the underlying biological processes \cite{eisen1998cluster,spellman1998}. Module network inference takes this principle one step further, and aims to infer simultaneously coexpression modules and their upstream regulators \cite{segal2003,friedman2004}. From a statistical perspective, modularity allows to reduce the number of  model parameters that need to be determined, because it is assumed that genes belonging to the same module share the same regulatory program, and therefore allows to learn more complex models, in particular non-linear probabilistic graphical models \cite{koller2009}, than would otherwise be possible.

While module networks were originally introduced to infer gene regulatory networks from gene expression data alone \cite{segal2003}, the method has meanwhile been extended to also include expression quantitative trait loci data \cite{lee2006, zhang2010bayesian}, regulatory prior data \cite{lee2009learning}, microRNA expression data \cite{bonnet2010a}, clinical data \cite{bonnet2010b}, copy number variation data \cite{akavia2010,bonnet2015} or protein interaction networks \cite{novershtern2011physical}. Furthermore, the method can be combined with gene-based network inference methods \cite{michoel2009b, roy2013integrated}. Finally, the module network method has been applied in numerous biological, biotechnological and biomedical studies \cite{segal2007,zhu2007b,li2007,novershtern2008,amit2009,vermeirssen2009,  novershtern2011densely, zhu2012reconstructing, arhondakis2016silico,behdani2017construction, marchi2017multidimensional}.

An area of interest that has received comparatively limited attention to date concerns the inference of \emph{differential} module networks. Differential networks extend the concept of differential expression, and are used to model how coexpression, regulatory or protein-protein interaction networks differ between two or more experimental conditions, cell or tissue types, or disease states \cite{de2010differential,ideker2012differential}. Existing differential network inference methods are mainly based on pairwise approaches, either by testing for significant differences between correlation values in different conditions, or by estimating a joint graph from multiple data sets simultaneously using penalized likelihood approaches \cite{gambardella2013differential, ha2015dingo, mckenzie2016dgca, voigt2017composite}. The inference of differential module networks is more challenging, because it requires a matching or comparable set of modules across the conditions of interest. A related problem has been addressed in a study of the evolutionary history of transcriptional modules in a complex phylogeny, using an algorithm that maps modules across species and allows to compare their gene assignments \cite{roy2013arboretum}.

In this chapter, we review the theoretical principles behind module network inference, explain practical protocols for learning module networks using the Lemon-Tree software \cite{bonnet2015}, and show in a concrete application on human gene expression data how the software can also be used to infer differential module networks using a similar principle as in \cite{roy2013arboretum}.

\section{Module network inference: theory and algorithms}
\label{sec:module-netw-infer}

\subsection{The module network model}
\label{sec:module-network-model}

Module networks are probabilistic graphical models \cite{friedman2004,koller2009} where each gene $g_i$, $i\in\{1,\dots,G\}$, is represented by a random variable $X_i$ taking continuous values. In a standard probabilistic graphical model or Bayesian network, it is assumed that the distribution of $X_i$ depends on the expression level of a set of regulators $\Pa_i$ (the ``parents'' of gene $i$). If the causal graph formed by drawing directed edges from parents to their targets is acyclic, then the joint probability distribution for the expression levels of all genes can be written as a product of conditional distributions, 
\begin{equation}\label{eq:7}
  p(x_1,\dots,x_G) = \prod_{i=1}^G p \bigl(x_i \mid \{x_j\colon 
  j\in\Pa_i\}\bigr).
\end{equation}
In data integration problems, we are often interested in explaining patterns in one data type (e.g.\ gene expression) by regulatory variables in another data type (e.g.\ transcription factor binding sites, single nucleotide or copy number variations, etc.). In this case, the causal graph is bipartite, and the acyclicity constraint is satisfied automatically.

In a module network, we assume that genes are partitioned into \emph{modules}, such that genes in the same module share the same parameters in the distribution function (\ref{eq:7}). Hence a module network is defined by a partition of $\{1,\dots,G\}$ into $K$ modules $\A_k$, a collection of parent genes $\Pa_k$ for each module $k$, and a joint probability distribution 
\begin{equation}\label{eq:2}
  p(x_1,\dots,x_G) = \prod_{k=1}^K \prod_{i\in\A_k} p 
  \bigl(x_i \mid \{x_j\colon   j\in\Pa_k\}\bigr).
\end{equation}
In a module network, only one conditional distribution needs to be parameterized per module, and hence it is clear that if $K\ll G$, the number of model parameters in eq.~\eqref{eq:2} is much smaller than in eq.~\eqref{eq:7}. Moreover, data from genes belonging to the same module are effectively pooled, leading to more robust estimates of these model parameters. This is the main benefit of the module network model.

\begin{figure}
  \centering
  \includegraphics[width=\linewidth]{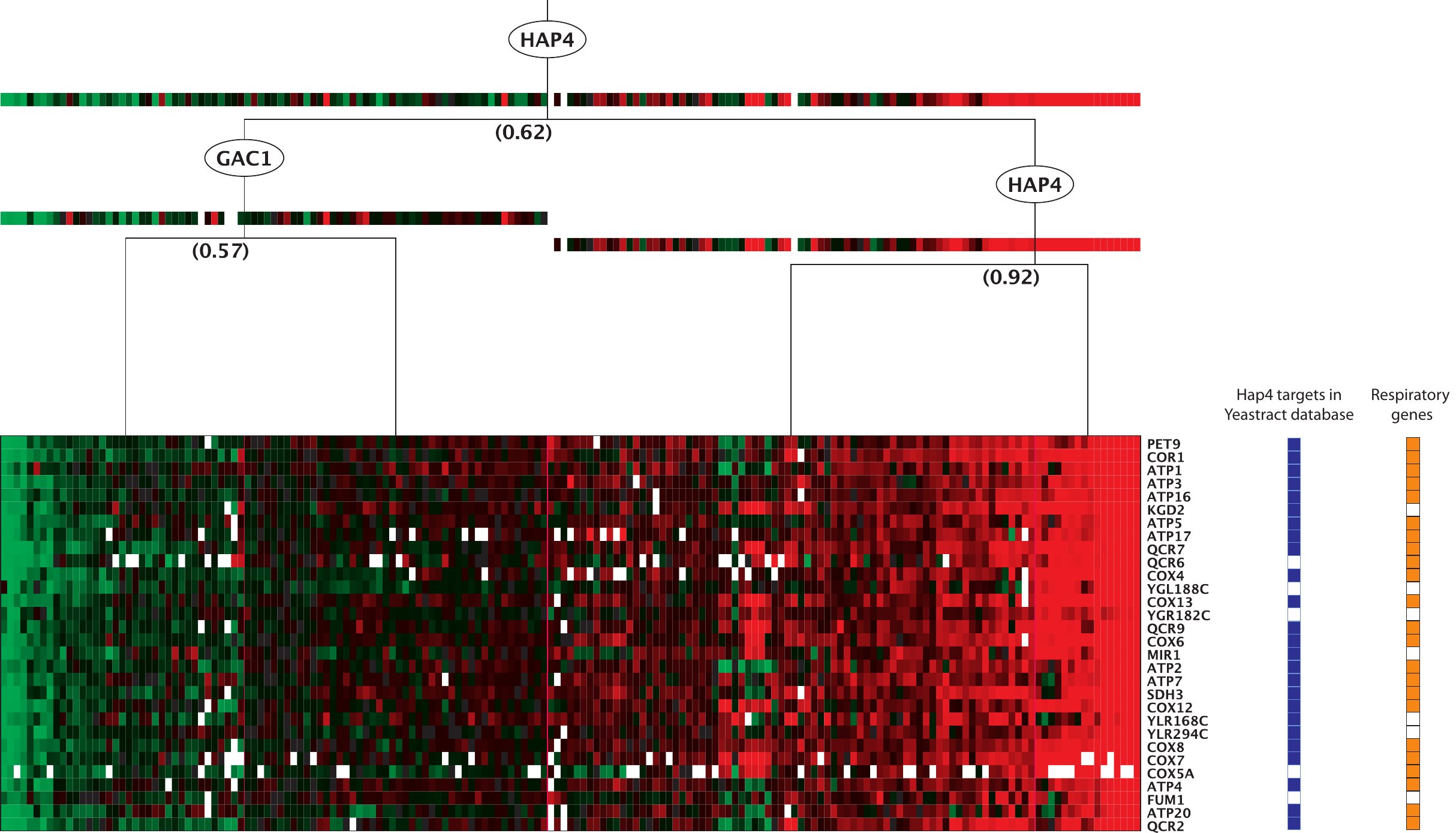}
  \caption{Example of a module and regulatory decision tree inferred from yeast data, with Hap4 assigned as a top regulator. Genes known to be regulated by Hap4 in YEASTRACT are marked in blue and those involved in respiration are marked in orange. Reused from\emph{ Joshi \emph{et al.}, Module networks revisited: computational assessment and prioritization of model predictions, Bioinformatics, 2009, 25(4):490--496} \cite{joshi2009}, by permission of Oxford University Press.}
  \label{fig:regr-tree}
\end{figure}

In principle, any type of conditional distribution can be used in eq.~\eqref{eq:2}. For instance, in a linear Gaussian framework \cite{koller2009}, one would assume that each gene is normally distributed around a linear combination of the parent expression levels. However, the pooling of genes into modules allows for more complex, non-linear models to be fitted. Hence it was proposed that the conditional distribution of the expression level of the genes in module $k$ is normal with mean and standard deviation depending on the expression values of the parents of the module through a regression tree (the \emph{``regulatory program''} of the module)  \cite{segal2003} (Figure~\ref{fig:regr-tree}). The tests on the internal nodes of the regression tree are usually defined to be of the form $x\geq v$ or not, for a split value $v$, where $x$ is the expression value of the parent associated to the node.

Given a module network specification $\M$, consisting of gene module assignments, regulatory decision trees, and normal distribution parameters at the leaf nodes, the probability density of observing an expression data matrix $\Dc=(x_{im})\in\R^{G\times N}$ for $G$ genes in $N$ samples is given by
\begin{equation*}
  P(\Dc\mid\M) = \prod_{m=1}^N \prod_{k=1}^K \prod_{i\in\A_k} p 
  \bigl(x_{im} \mid \{x_{jm}\colon   j\in\Pa_k\}\bigr) = \prod_{k=1}^K \prod_{\ell=1}^{L_k} \prod_{i\in\A_k} \prod_{m\in \Ec_\ell} p(x_{im}\mid \mu_\ell,\sigma_\ell),
\end{equation*}
where $L_k$ is the number of leaf nodes of module $k$'s regression tree, $\Ec_\ell$ denotes the experiments that end up at leaf $\ell$ after traversing the regression tree, and $(\mu_\ell,\sigma_\ell)$ are the normal distribution paramaters at leaf $\ell$. The Bayesian model score is obtained by taking the log-marginal probability over the parameters of the normal distributions at the leaves of the regression trees with a normal-gamma prior:
\begin{align}
  S &= \sum_k S_k = \sum_k\sum_\ell S_k(\Ec_\ell) \label{eq:S1}\\
  S_k(\Ec_\ell) &= -\tfrac12 R_0^{(\ell)}\log(2\pi) + \tfrac12
  \log\bigl(\frac{\lambda_0}{\lambda_0 + R_0^{(\ell)}}\bigr) 
   - \log\Gamma(\alpha_0) + \log\Gamma(\alpha_0
  + \tfrac12 R_0^{(\ell)}) \nonumber\\
  &\qquad + \alpha_0\log\beta_0 -(\alpha_0 + \tfrac12
  R_0^{(\ell)})\log\beta_1 \nonumber
\end{align}
where $ R_q^{(\ell)}$ are  the sufficient statistics at leaf $\ell$,
\begin{equation*}
  R_q^{(\ell)} = \sum_{m\in\Ec_\ell} \sum_{i\in\A_k} x_{i,m}^q\,,\; q=0,1,2,
\end{equation*}
and
\begin{equation*}
  \beta_1 = \beta_0 + \frac12\Bigl[ R_2^{(\ell)} -
  \frac{(R_1^{(\ell)})^2}{R_0^{(\ell)}} \Bigr] 
  + \frac{\lambda_0 \bigl( R_1^{(\ell)} - \mu_0 R_0^{(\ell)}
    \bigr)^2}{2(\lambda_0 + R_0^{(\ell)})R_0^{(\ell)}}.
\end{equation*}
Details of this calculation can be found in \cite{segal2005,michoel2007a}.

\subsection{Optimization algorithms}
\label{sec:optim-algor}

The first optimization strategy proposed to identify high-scoring module networks was a \textbf{greedy hill-climbing} algorithm \cite{segal2003}. This algorithm starts from an initial assignment of genes to coexpression clusters (e.g. using k-means), followed by assigning a new regulator to each module by iteratively finding the best (if any) new split of a current leaf node into two new leaf nodes given the current set of gene-to-module assignments, and reassigning genes between modules given the current regression tree, while preserving acyclicity throughout. The decomposition of the Bayesian score [eq.~\eqref{eq:S1}] as a sum of leaf scores of the different modules allows for efficient updating after every regulator addition or gene reassignment.

An improvement to this algorithm was found, based on the observation that the Bayesian score depends only on the assignment of samples to leaf nodes, and not on the actual regulators or tree structure that induce this assignment \cite{michoel2007a}. Hence, a \textbf{decoupled greedy hill-climbing} algorithm was developed, where first the Bayesian score is optimized by two-way clustering of genes into modules and samples into leaves for each module, and then a regression tree is found for the converged set of modules by hierarchically merging the leave nodes and finding the best regulator to explain the split below the current merge. This algorithm achieved comparable score values as the original one, while being considerably faster \cite{michoel2007a}.

Further analysis of the greedy two-way clustering algorithm revealed the existence of multiple local optima, in particular for moderate to large data sets ($\sim$1000 genes or more), where considerably different module assignments result in near-identical scores. To address this issue, a \textbf{Gibbs sampler} method was developed, based on the Chinese restaurant process \cite{qin2006}, for sampling from the posterior distribution of two-way gene/sample clustering solutions \cite{joshi2008}. By sampling an ensemble of multiple, equally probable solutions, and extracting a core set of `tight clusters' (groups of genes which consistenly cluster together), gene modules are identified that are more robust to fluctuations in the data and have higher functional enrichment compared to the greedy clustering strategies \cite{joshi2008,joshi2009}.

Finally, the Gibbs sampling strategy for module identification was complemented with a probabilistic algorithm, based on a logistic regression of sample splits on candidate regulator expression levels,  for sampling and ensemble averaging of regulatory programs, which resulted in more accurate regulator assignments \cite{joshi2009}.

\section{The Lemon-Tree software suite for module network inference}
\label{sec:lemon-tree-software}

\subsection{Lemon-Tree software package}

Lemon-Tree is a software suite implementing all of the algorithms discussed in Section \ref{sec:optim-algor}. Lemon-Tree has been benchmarked using large-scale tumor datasets and shown to compare favorably with other module network inference methods \cite{bonnet2015}. Its performance has been carefully assessed also in an independent study not involving the software authors \cite{lu2017dissection}. Lemon-Tree is self-contained, with no external program dependencies, and is entirely coded in the Java\textsuperscript{TM} programming language. Users can download a pre-compiled version of the software, or alternatively they can download and compile the software from the source code, which is available on the GitHub repository (\url{https://github.com/eb00/lemon-tree}). Note that there is also a complete wiki on the Lemon-Tree GitHub (\url{https://github.com/eb00/lemon-tree/wiki}), with detailed instruction on how to download, compile, use the software, what are the default parameters and an extended bibliography on the topic of module networks.

Lemon-Tree is a command-line software, with no associated graphical user interface at the moment. The different steps for building the module network are done by launching commands with different input files that will generate different output files. All the command line examples below are taken from the Lemon-Tree tutorial, that users are encouraged to download and reproduce by themselves.

\begin{figure}
  \centering
  \includegraphics[width=\linewidth]{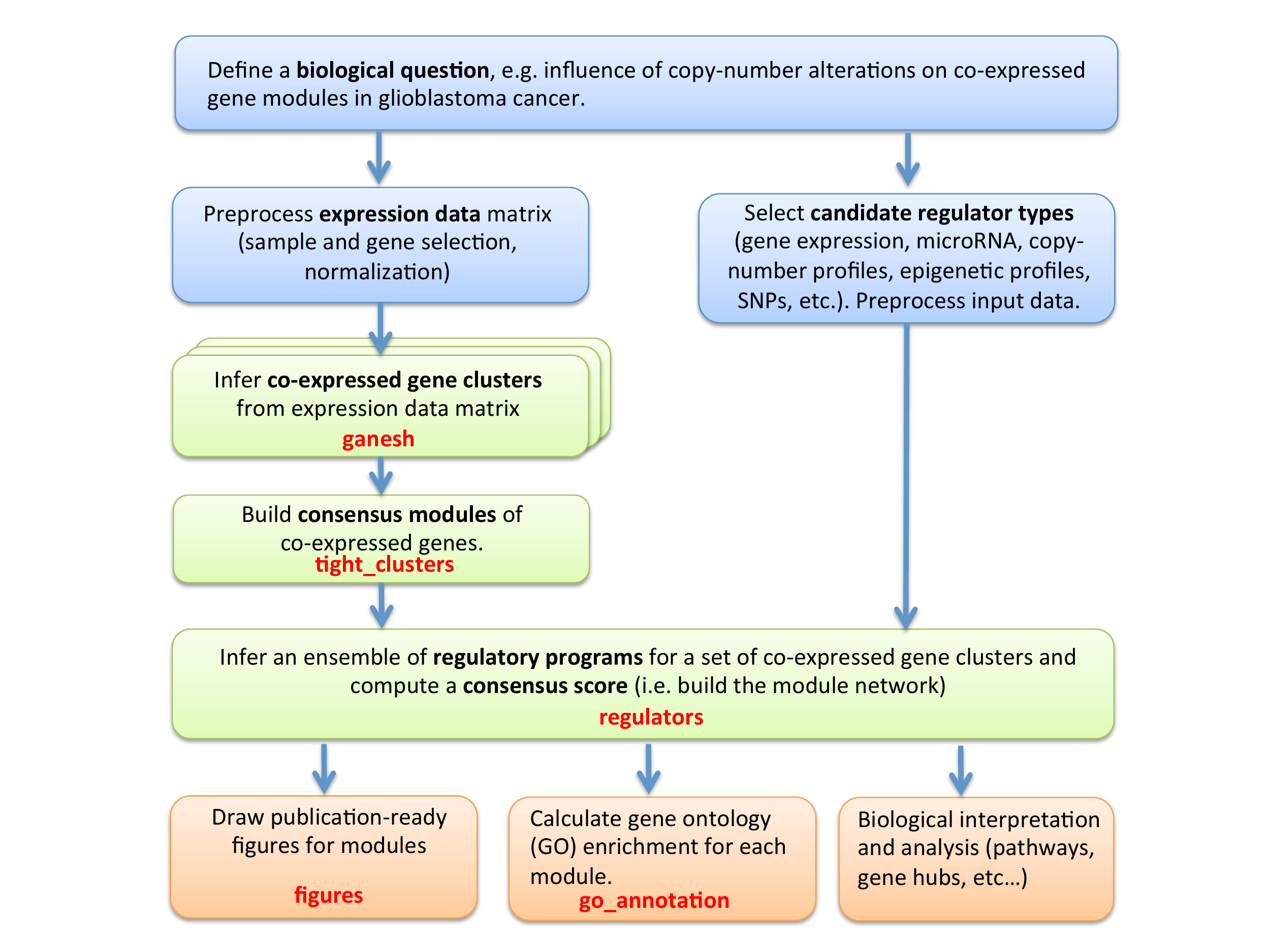}
  \caption{{\bf Flow chart for module network inference
    with Lemon-Tree.}  This figure shows the general workflow for a typical integrative module network inference with Lemon-Tree. Blue  boxes indicate the pre-processing steps that are done using  third-party software such as R or user-defined scripts. Green boxes  indicates the core module network inference steps done with the  Lemon-Tree software package.  Typical post-processing tasks (orange  boxes), such as GO enrichment calculations, can be performed with  Lemon-Tree or other tools. The Lemon-Tree task names are indicated  in red (see main text for more details). Figure reproduced from \cite{bonnet2015} under Creative Commons Attribution License.}
  \label{fig:flow}
\end{figure}

The purpose of the Lemon-Tree software package is to create a module network from different types of 'omics' data. The end result is a set of gene clusters (co-expressed genes), and their associated ``regulators''. The regulators can be of different types, for instance mRNA expression, copy-number profiles, variants (such as single nucleotide variants) or even clinical parameter profiles can be used. There are three fundamental steps or tasks to build a module network with Lemon-Tree (Figure~\ref{fig:flow}):
\begin{itemize}
\item Generate several cluster solutions ("ganesh" task).
\item Merge the different cluster solutions using the fuzzy clustering algorithm ("tight\_clusters" task).
\item Assign regulators to each cluster, producing the module network ("regulators" task).
\end{itemize}
    
\subsection{Ganesh task}
\label{sec:ganesh-task}

The goal of this task is to cluster genes from a matrix (rows) using a probabilistic algorithm (Gibbs sampling) \cite{joshi2008}. This step is usually done on the mRNA expression data only, although some other data type could be used, for instance proteomic expression profiles. We first select genes having non-flat profiles, by keeping genes having a standard deviation above a certain value (0.5 is often used as the cutoff score, but this value might depend on the dataset). The data is then centered and scaled (by row) to have a mean of 0 and a standard deviation of 1. To find one clustering solution, the following command can be used (the command is spread here over multiple lines, but should be entered on a single line without the backslash characters):

\begin{verbatim}
java -jar lemontree.jar -task ganesh \
-data_file data/expr_matrix.txt \
-output_file ganesh_results/cluster1
\end{verbatim}

The clustering procedure should be repeated multiple times, using the same command, only changing the name of the output file. For instance we could generate 5 runs, named cluster1, cluster2, cluster3, cluster4 and cluster5, with the same command, just by changing the name of the output file.

\subsection{Tight clusters task}
\label{sec:tight-clusters-task}

Here, we are going to generate a single, robust clustering solution from all the individual solutions generated at the previous step, using a graph clustering algorithm \cite{michoel2012}. Basically, we group together genes that frequently co-occur in all the solutions. Genes that are not strongly associated to a given cluster will be eliminated.

\begin{verbatim}
java -jar lemontree.jar -task tight_clusters \
-data_file data/expr_matrix.txt \
-cluster_file cluster_file_list \
-output_file tight_clusters.txt \
-node_clustering true
\end{verbatim}

The ``cluster\_file'' is a simple text file, listing the location of all the individual cluster files generated at the ``ganesh'' step. By default, the tight clusters procedure is keeping only clusters that have a minimum of 10 genes (this can be easily changed by overriding a parameter in the command).

\subsection{Revamp task}
\label{sec:revamp-task}

This task is aimed at maximizing the Bayesian coexpression clustering score of an existing module network while preserving the initial number of clusters. A threshold can be specified to avoid that genes are reassigned if the score gain is below this threshold and allowing the systematic tracking of the conservation and divergence of modules with respect to the initial partition. This task can be used to optimize an existing module network obtained with a different clustering algorithm, or to optimize an existing module network for a different data matrix, e.g. a subset of samples as presented in Section \ref{sec:diff-module-netw}.

\begin{verbatim}
java -jar lemontree.jar -task revamp \
-data_file data/expr_matrix.txt \
-cluster_file cluster_file.txt \
-reassign_thr 0.0 \
-output_file revamped_clusters.txt \
-node_clustering true
\end{verbatim}

The ``cluster\_file.txt'' is a simple text clustering file, like the one obtained in ``tight\_clusters" step, and ``reassign\_thr'' is the score gain threshold that must be reached to move a gene from one cluster to another. By default, this reassignment threshold is set to 0.

\subsection{Regulators task}
\label{sec:regulators-task}

In this task, we assign sets of ``regulators'' to each of the modules using a probabilistic scoring, taking into account the profile of the candidate regulator and how well it matches the profiles of co-expressed genes \cite{joshi2009}. The candidate regulators can be divided in two different types, depending on the nature of their profiles: continuous or discrete. The first type can be for example transcription factors or signal transducers mRNA expression profiles (selected from the same matrix used for detecting co-expressed genes), microRNA expression profiles or gene copy-number variants profiles (CNVs). For the latter, the  numerical values will be integers, such as the different clinical grades characterizing a disease state (discrete values), or single nucleotides variants profiles (SNVs, characterized by profiles with 0/1 values). In all cases, the candidate regulator profiles must have been made on the same samples as the tight clusters defined previously. Missing values are allowed, but obviously they should not constitute the majority of the values in the profile. Note that a patch to the regulator assignment implementation identified in \cite{lu2017dissection} is included in Lemon-Tree version 3.0.5 or above.

Once the list of candidate regulators is established, the assignment to the clusters can be made with a single command like this:

\begin{verbatim}
java -jar lemontree.jar -task regulators 
-data_file data/expr_matrix.txt \
-reg_file data/reg_list.txt \
-cluster_file tight_clusters.txt \
-output_file results/reg_tf
\end{verbatim}

The ``reg\_file'' option is a simple text list of candidate regulators that are present in the expression matrix. If the regulators are discrete, it is mandatory to add a second column in the text file, describing the type of the regulator (``c'' for continuous or ``d'' for discrete). The profiles for co-expressed genes and for all the regulators must be included in the matrix indicated by the data\_file parameter. 

Note that this command will create four different output files, using the ``output\_file'' parameter as the prefix for all the files.

\begin{itemize}
\item    reg\_tf.topreg.txt: Top 1\% regulators assigned to the modules.
\item    reg\_tf.allreg.txt: All the regulators assigned.
\item   reg\_tf.randomreg.txt: Regulators assigned randomly to the modules.
\item   reg\_tf.xml.gz: xml file containing all the regulatory trees used for assigning the regulators.
\end{itemize}

The regulators text files all have the same format: three columns representing respectively the regulator name, the module number and the score value.

\subsection{Figures task}
\label{sec:figures-task}

This task is creating one figure per module. The figure represent the expression values color-coded with a gradient ranging from dark blue (low expression values) to bright yellow (high expression values). All the module genes are in the lower panel while the top regulators for the different classes or types of regulators (if any) are displayed in the upper panel. A regulation trees is represented on top of the figure, with the different split points highlighted on the figure as vertical red lines. The name of each gene is displayed on the left of the figure.

\begin{verbatim}
java -jar lemontree.jar \
-task figures \
-top_regulators reg_files.txt \
-data_file data/all.txt \
-reg_file data/reg_list.txt \
-cluster_file tight_clusters.txt \
-tree_file results/reg_tf.xml.gz
\end{verbatim}

Note that the ``top\_regulators'' parameter is a simple text file listing the different top regulator files and their associated clusters. Such a file could be for instance the file reg\_tf.topreg.txt mentionned in the previous paragraph. All figures are generated to the eps (encapsulated postcript) format, but it is relatively easy to convert this format to other common formats such as pdf.

\subsection{GO annotation task}
\label{sec:go-annotation-task}

The goal of this task is to calculate the GO (Gene Ontology) category enrichment for each module, using code from the BiNGO package \cite{maer05b}. We have to specify two GO annotation files that are describing the GO codes associated with the genes (``gene\_association.goa\_human'') and another file describing the GO graph (``gene\_ontology\_ext.obo''). These files can be downloaded for various organisms from the GO website (\url{http://www.geneontology.org}). We also specify the set of genes that should be used as the reference for the calculation of the statistics, in this case the list of all the genes that are present on the microarray chip (file ``all\_gene\_list''). The results are stored in the output file ``go.txt''.

\begin{verbatim}
java -jar lemontree.jar \
-task go_annotation \
-cluster_file tight_clusters.txt \
-go_annot_file gene_association.goa_human \
-go_ontology_file gene_ontology_ext.obo \
-go_ref_file all_gene_list \
-output_file go.txt
\end{verbatim}

\section{Differential module network inference}
\label{sec:diff-module-netw}

\subsection{Differential module network model}
\label{sec:diff-module-netw-1}

Assume that we have expression data in $T$ different conditions (e.g., experimental treatments, cell or tissue types, disease stages or states), with $N_t$ samples in each condition $t\in\{1,\dots,T\}$, and wish to study how the gene regulatory network differs (or not) between conditions. We define a differential module network as a collection of module networks $\{\M_1,\dots,\M_T\}$, one for each condition, subject to constraints, and model gene expression levels for $G$ genes as 
\begin{equation}\label{eq:1}
  p\bigl(x_1,\dots, x_G \mid \M_1,\dots,\M_T \bigr) = \prod_{t=1}^T p\bigl(x_1,\dots, x_G \mid \M_t\bigr),
\end{equation}
where each factor is a model of the form of eq.~\eqref{eq:2}. Hence, if gene $i$ is assigned to modules $\{k_{i1},\dots,k_{iT}\}$ in each module network, its parent set is the union $\Pa_i=\cup_{t=1}^T \Pa_{k_{it}}$. If the graph mapping these parent sets to their targets is acyclic, eq.~\eqref{eq:1} defines a proper Bayesian network. If the the individual factors $p\bigl(x_1,\dots, x_G \mid \M_t\bigr)$ are the usual Gaussians with parameters depending on the parent expression levels in that module network, their product remains a Gaussian. By Bayes' theorem we can write, for a concatenated data matrix $\Dc = (\Dc_1,\dots,\Dc_T)$,
\begin{equation}\label{eq:3}
  p\bigl( \M_1,\dots,\M_T  \mid \Dc\bigr) \propto  p\bigl( \M_1,\dots,\M_T\bigr) \prod_{t=1}^T p\bigl( \Dc_t \mid \M_t\bigr) 
\end{equation}
If we assume independence, $p(\M_1,\dots,\M_T)=\prod_t p(\M_t)$, then optimization of, or  sampling from, eq.~\eqref{eq:3}, is the same as inferring module networks independently in each condition, but this will reveal little of the underlying relations between the conditions. Instead we assume that there exists a conserved set of modules across all conditions, but their gene and regulator assignment may differ in each condition. This results in the following constraints:
\begin{enumerate}
\item The number of modules must be the same in each module network, i.e.\\ $p(\M_1,\dots,\M_T)=0$ unless $K_1=\dots=K_T=K$.
\item Module networks with more similar gene and/or  regulator assignments are more likely \textit{a priori},
  \begin{equation}\label{eq:4}
    \log p\bigl(\M_1,\dots,\M_T\bigr) = -\sum_{k=1}^K\sum_{t,t'} \Bigl[ \lambda_{t,t'} f \bigl(\A_k^{(t)},\A_k^{(t')}\bigr) + \mu_{t,t'} g\bigl(\Pa_k^{(t)},\Pa_k^{(t')}\bigr) \Bigr],
  \end{equation}
where $f$ and $g$ are distance functions on sets (e.g.\ Jaccard distance) and $\lambda_{t,t'}$ and $\mu_{t,t'}$ are penalty parameters that encode the relative \textit{a priori} similarity between conditions.
\end{enumerate}

\subsection{Optimization algorithm}
\label{sec:optim-algor-1}

For simplicity we assume here that $\mu_{t,t'}=0$ and $\lambda_{t,t'}=\lambda$ for all $(t,t')$ in eq.~\eqref{eq:4}, i.e. we will only constrain the gene assignments, and uniformly so for all condition pairs; the complete model will be treated in detail elsewhere. More general forms of $\lambda_{t,t'}$ can be used for instance to mimic the model of \cite{roy2013arboretum}, where conditions represented different species and gene reassignments were constrained by a phylogenetic tree. With a fixed $\lambda$, instead of modelling $\lambda$ and $f$ explicitly, we observe that the effect of including $f$ in the model \eqref{eq:3} is to impose a penalty on gene reassignments: starting from identical modules in all conditions, a gene reassignment in condition $t$ increases the posterior log-likelihood only if its increase in $\log p(\Dc_t\mid \M_t)$ is sufficiently large to overcome the penalty induced by eq.~\eqref{eq:4}. This can be modelled equivalently by setting a uniform module reassignment score threshold as an external parameter. Hence we propose the following heuristic optimization algorithm for differential module network inference using Lemon-Tree:
\begin{enumerate}
\item Create a concatenated gene expression matrix $\Dc = (\Dc_1,\dots,\Dc_T)$ and learn a set of coexpression modules using tasks ``ganesh'' (Section \ref{sec:ganesh-task} and ``tight clusters'' (Section \ref{sec:tight-clusters-task}). This results in a set of module networks $(\M_1,\dots,\M_T)$ with identical module assignments and empty parent sets.
\item Set a reassignment threshold value and use task ``revamp'' (Section \ref{sec:revamp-task}) to maximize the Bayesian coexpression clustering score $\log p(\Dc_t\mid \M_t)$ [cf. eq.~\eqref{eq:S1}] for each condition independently, but subject to the constraint that gene reassignments must pass the Bayesian score difference threshold.
\item Assign regulators to each module for each condition independently using task ``regulators'' (Section \ref{sec:regulators-task}).
\end{enumerate}

\subsection{Reconstruction of a differential module network between atherosclerotic and non-atherosclerotic arteries in cardiovascular disease patients}
\label{sec:reconstr-diff-module}

To illustrate the differential module network inference algorithm, we applied it to 68 atherosclerotic (i.e. diseased) arterial wall (AAW) samples and 79 non-atherosclerotic (i.e. non-diseased) internal mammary artery (IMA) samples from the Stockholm Atherosclerosis Gene Expression study \cite{hagg2009multi,foroughi2015,talukdar2016}, using 1803 genes with variance greater than 0.5 in the concatenated data. The STAGE study was designed to study the effect of genetic variation on tissue-specific gene expression in cardiovascular disease \cite{foroughi2015}. According to the systems genetics paradigm, genetic variants in regulatory regions affect nearby gene expression (``\textit{cis}-eQTL effects''), which then causes variation in downstream gene networks (``\textit{trans}-eQTL effects'') and clinical phenotypes \cite{schadt2009,talukdar2016}. We therefore considered as candidate regulators the tissue-specific sets of genes with significant eQTLs \cite{foroughi2015} and present in our filtered gene list (668 AAW and 964 IMA genes, 267 in common), and ran the ``regulators'' task on each set of modules independently. 

\begin{figure}
  \centering
    \includegraphics[width=0.49\linewidth]{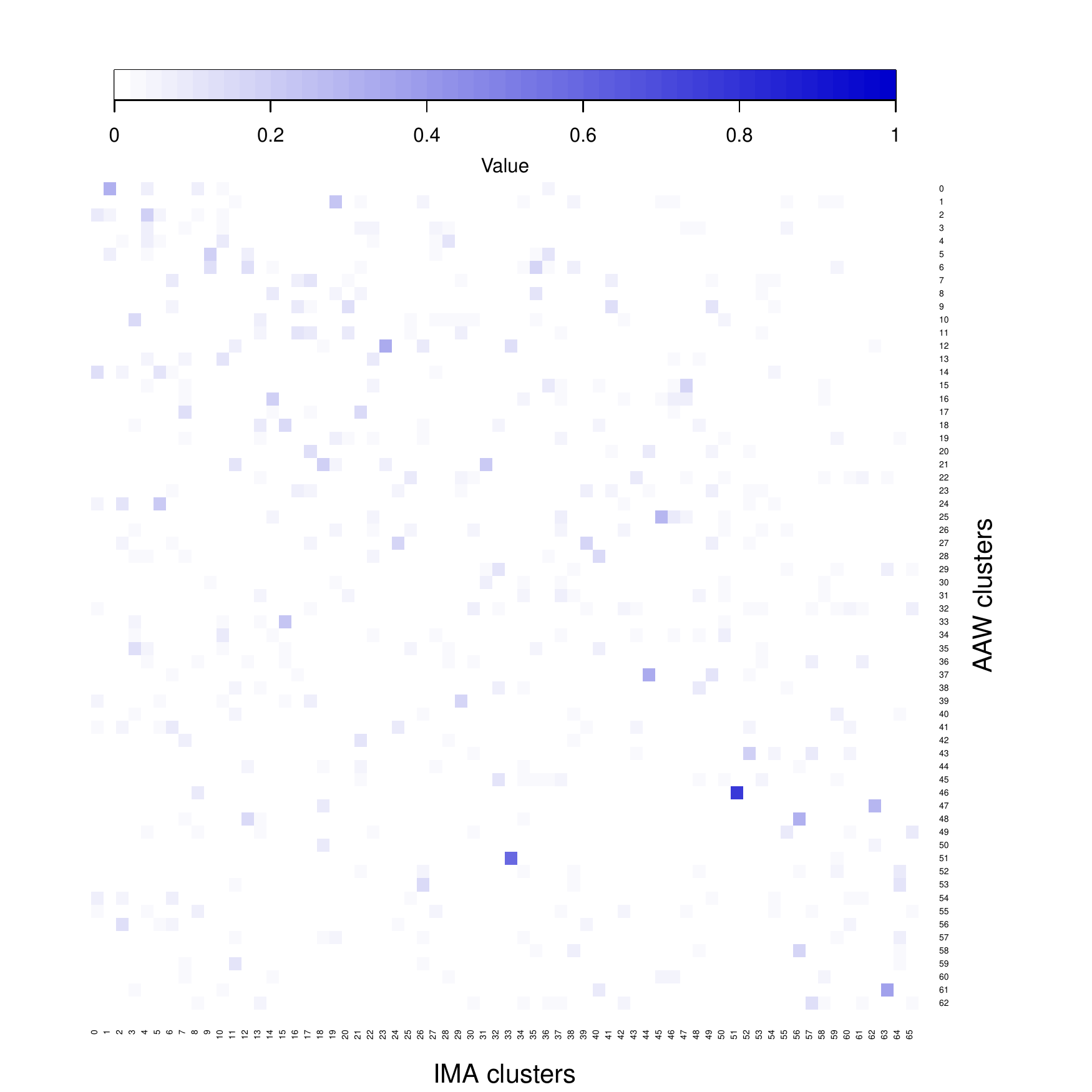}
    \includegraphics[width=0.49\linewidth]{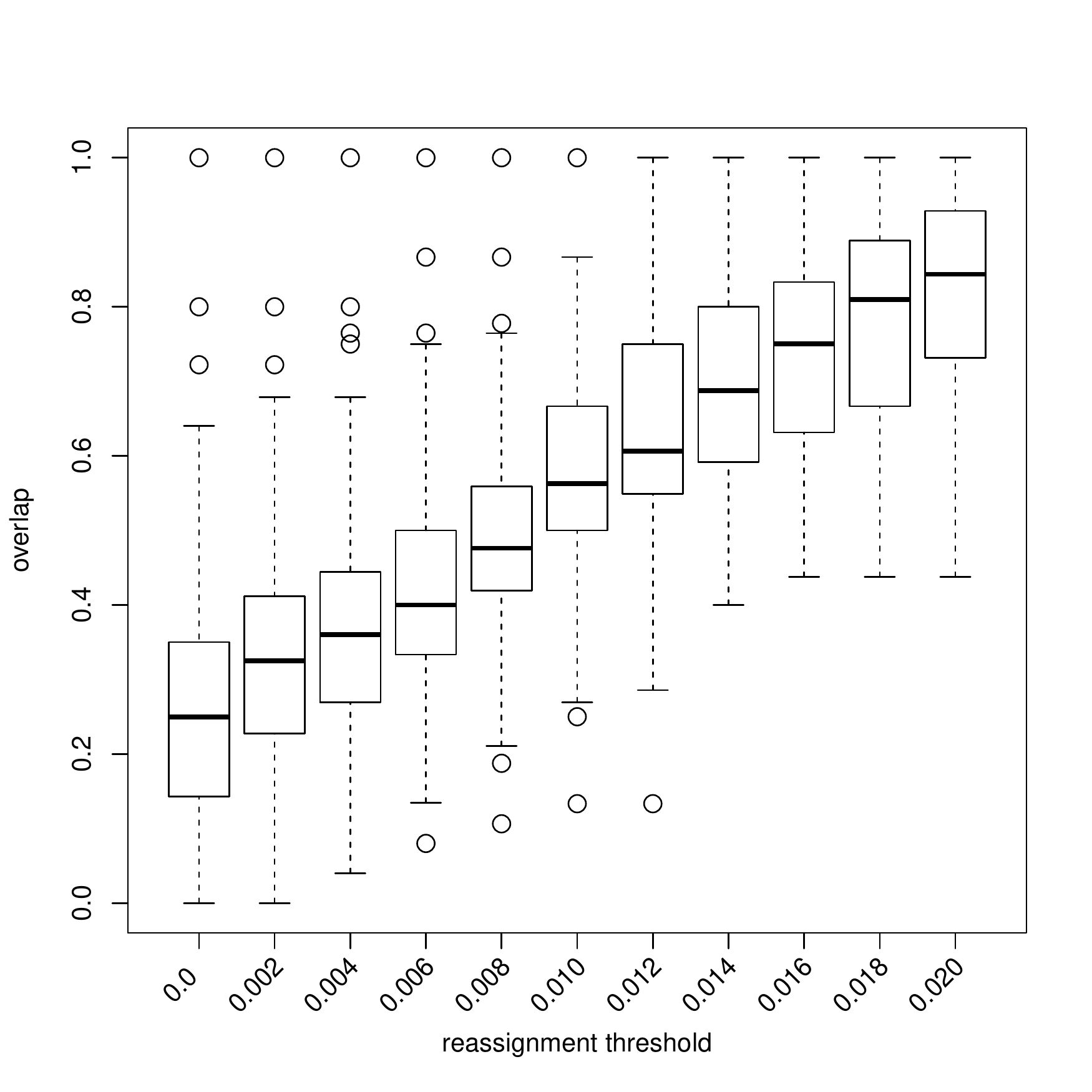}
  \caption{\textbf{Differential module network inference on STAGE AAW and IMA tissues.} \textbf{(a)} Independent clustering of tissue-specific data results in poorly identifiable module relations between tissues. Shown is the pairwise overlap fraction for all pairs of modules inferred in AAW (rows) and IMA (columns). \textbf{(b)} Joint clustering of data across both tissues using the ``revamp'' task in Lemon-Tree results in a one-to-one mapping of modules with a tunable level of overlap. Shown are the module overlap distributions (boxplots) at different values for the tuning parameter.}
  \label{fig:diff-modnet}
\end{figure}

As expected, independent clustering of the two data sets results in different numbers of modules, and an inability to map modules unambiguously across tissues (Figure~\ref{fig:diff-modnet}a). In contrast, application of the differential module network optimization algorithm (Section \ref{sec:optim-algor-1}) results in a one-to-one mapping of modules, whose average overlap varies smoothly as a function of the reassignment threshold value (Figure~\ref{fig:diff-modnet}b). 

\begin{figure}
  \centering
  \includegraphics[width=\linewidth]{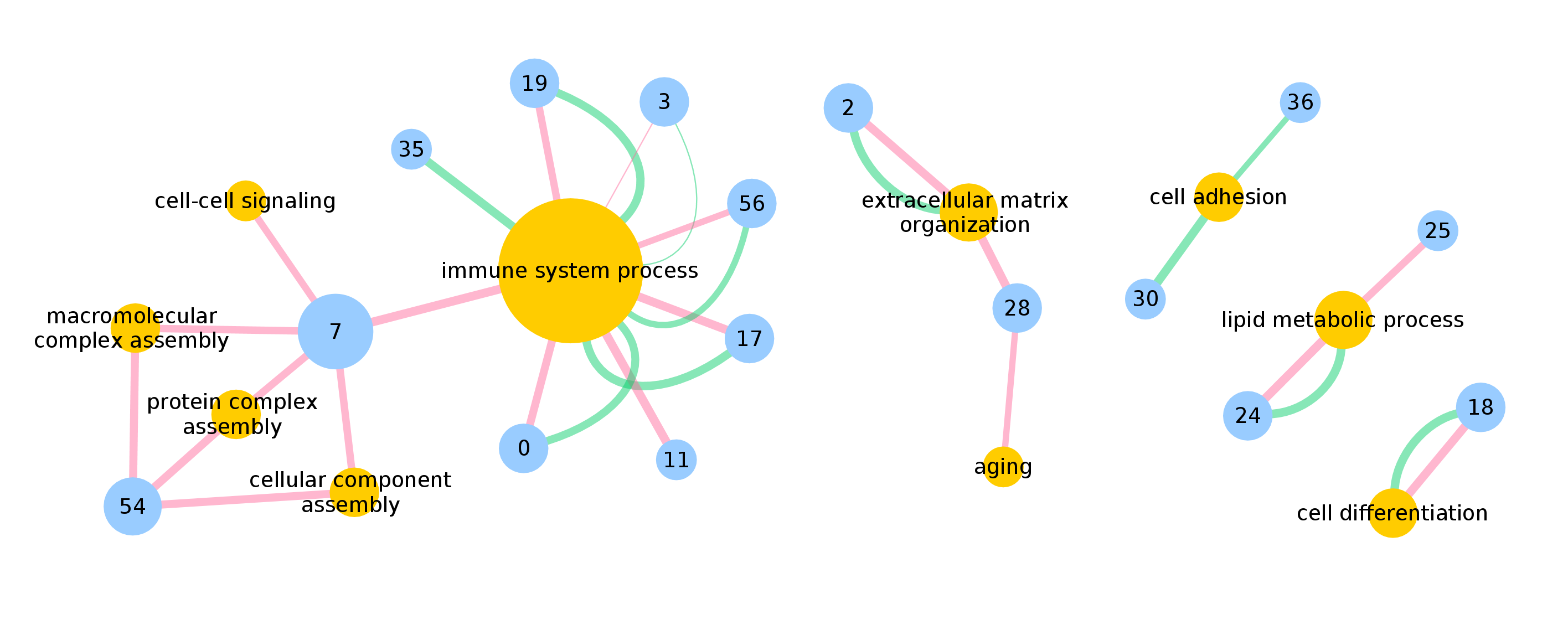}

  \caption{\textbf{Enrichment for GO Slim terms in the STAGE AAW-IMA differential module network.} Blue nodes are modules, yellow nodes GO Slim terms. Red and green edges indicate enrichment ($q<0.05$) in the corresponding AAW and IMA module, respectively. The reassignment threshold used is 0.015.}
  \label{fig:go-slim}
\end{figure}

The biological assumption underpinning the differential module network model (Section \ref{sec:diff-module-netw-1}) is that each module represents a higher-level biological process, or set of processes, that is shared between conditions, whereas the differences in gene assignments reflect differences in molecular pathways that are affected by, or interact with, this higher-level process. To test whether the optimization algorithm accurately captures this biological picture, we first performed gene ontology enrichment (task ``go\_annotation'', Section \ref{sec:go-annotation-task}) using the GO Slim ontology. GO Slims give a broad overview of the ontology content without the detail of the specific fine-grained terms (\url{http://www.geneontology.org/page/go-slim-and-subset-guide}). Consistent with our biological assumption, matching modules in atherosclerotic and non-atherosclerotic tissue are often enriched for the same GO Slim categories (Figure~\ref{fig:go-slim}).

\begin{figure}
  \centering
    \includegraphics[width=1.05\linewidth]{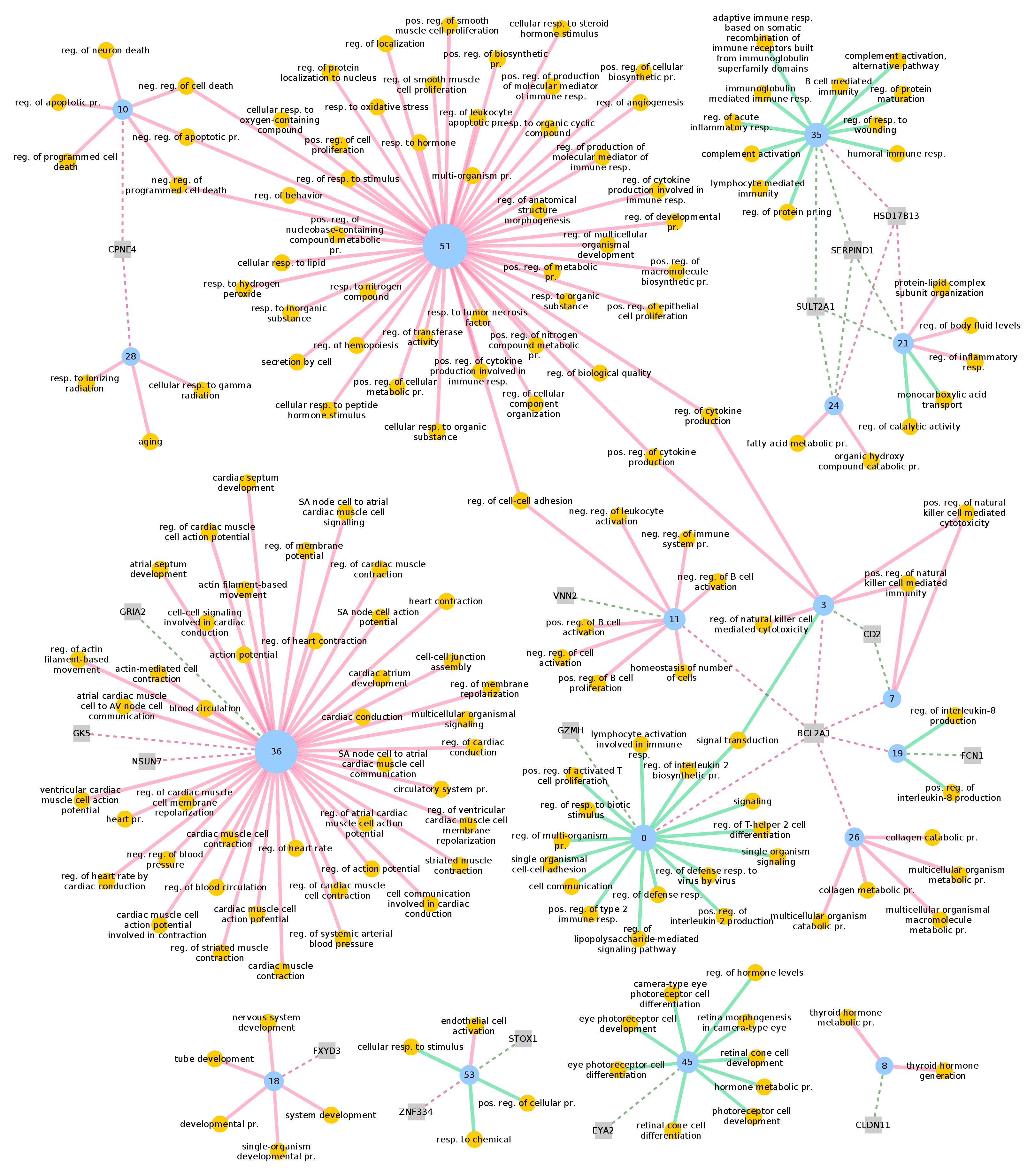}

  \caption{\textbf{Tissue-specific GO enrichment for terms related to the immune system process (GO0002376) and regulator assignment in the STAGE AAW-IMA differential module network.} Reassignment of nodes was computed with a threshold of 0.015. Blue nodes are modules, yellow nodes GO terms, grey nodes regulatory genes. Red and green edges indicate tissue-specific enrichment ($q<0.01$) in the corresponding AAW and IMA module, respectively. Dashed red and green edges indicate regulator assignments in AAW and IMA, respectively. Only top 1\% regulators are depicted. neg., negative; pos., positive; pr., process; reg., regulation; resp., response.}
  \label{fig:go-reg}
\end{figure}

Next, we performed gene ontology enrichment using the complete, fine-grained ontology, and removed all enrichments that were shared between matching modules. The resulting tissue-specific module enrichments reflected biologically meaningful differences between healthy and diseased arteries (Figure~\ref{fig:go-reg}). For instance, clusters 3 and 7 present a strong enrichment in AAW for the regulation of natural killer (NK) cells that augment atherosclerosis by cytotoxic-dependent mechanisms \cite{selathurai2014natural}. In IMA, these clusters are predicted to be regulated by genetic variation in CD2, a cell adhesian molecule found on the surface of T and NK cells, whereas in AAW their predicted regulator is BCL2A1, an important cell death regulator and pro-inflammatory gene that is upregulated in coronary plaques compared to healthy controls \cite{sikorski2014data}. This suggests that misregulation of cytotoxic response processes plays a role in the disease, further supported by the overrepresentation in cluster 10 of genes associated with cell death that are a important trigger of plaque rupture \cite{martinet2011pharmacological}.
Furthermore, variations in BCL2A1 are predicted to regulate other clusters exclusively in AAW too, with disease-relevant AAW enrichments. Cluster 11 is associated with the regulation of B lymphocytes, which may attenuate the neointimal formation of atherosclerosis \cite{gjurich2014selectin}, while cluster 26 is enriched for collagen production regulation. Uncontrolled collagen accumulation leads to arterial stenosis, while excessive collagen breakdown combined with inadequate synthesis weakens plaques thereby making them prone to rupture \cite{rekhter1999collagen}. Last, as expected, terms related with the heart, cardiac muscle and blood circulation are strongly enriched in AAW, in particular in cluster 36. In AAW, this cluster is regulated by GK5, which plays an important role in fatty acid metabolism  and whose upregulation has previously been associated to the pathogenesis of atherosclerosis and cardiovascular disease in patients with auto-immune conditions \cite{perez2015gene}. On the opposite side, cluster 36 in IMA is regulated by GRIA2, a player in the ion transport pathway, which has been shown to be down-regulated in advanced atherosclerotic lesions \cite{fu2008peripheral}.

In summary, this application has shown that differential module network inference allows to identify sets of one-to-one mapping modules representing broad biological processes conserved between conditions, with biologically relevant differences in  fine-grained  gene-to-module assignments and upstream regulatory factors.

\begin{acknowledgement}
PE and TM are supported by Roslin Institute Strategic Programme funding from the BBSRC [BB/P013732/1].
\end{acknowledgement}


\end{document}